\documentclass[aps,12pt,superscriptaddress,amsfonts,amssymb,amsmath]{revtex4-2}
\pdfoutput=1

\usepackage{graphicx}
\usepackage{epsfig}
\usepackage{makeidx}
\usepackage{epstopdf}
\usepackage{natbib}
\usepackage{xcolor}
\usepackage{amsmath}
\usepackage{bm}

\usepackage{verbatim}
\usepackage{subfig}
\usepackage{empheq}
\usepackage{dsfont}
\usepackage{hyperref}
\usepackage{caption}
\usepackage{url}

\newcommand{\angstrom}{\mbox{\normalfont\AA}}

\begin{document}

\title{Numerical 
simulations Unveil Superradiant Coherence in a Lattice of Charged Quantum Oscillators}

\author{L.\ Gamberale \footnote{Email address: luca.gamberale@mib.infn.it}}
\affiliation{Quantumatter Inc., Dover DE USA\\ LEDA srl, Universit\`a Milano Bicocca \\ I-20126 Milano, Italy}

\author{G.\ Modanese \footnote{Email address: giovanni.modanese@unibz.it}}
\affiliation{Free University of Bozen-Bolzano \\ Faculty of Engineering \\ I-39100 Bolzano, Italy}
\date{\today}

\linespread{0.9}

\begin{abstract}
A system of ${N_{osc}}$ charged oscillators interacting with the electromagnetic field, spatially confined in a 3D lattice of sub-wavelength dimension, can condense into a superradiant coherent state if appropriate density and frequency conditions are met. In this state, the common frequency $\omega$ of the oscillators and the plasma frequency $\omega_p$ of the charges are combined into a frequency $\omega'=\sqrt{\omega^2+\omega_p^2}$ that is off-shell with respect to the wavelength of the photon modes involved, preventing them from propagating outside the material. Unlike other atomic cavity systems, the frequency $\omega$ in this case is not determined by the cavity itself but is defined by the periodic electrostatic potential that confines the charged particles in the lattice. Additionally, the electromagnetic modes involved have wave vectors distributed
in all spatial directions, resulting in a significant increase in coupling. 
The analytical study of this system can be carried out in the limit of large ${N_{osc}}$ by searching for an approximation of the ground state via suitable coherent trial states. Alternatively, numerical simulations can be employed for smaller ${N_{osc}}$. In the numerical approach, it is possible to go beyond the Rotating Wave Approximation (RWA) and introduce a dissipation term for the photon modes. This dissipation term can account for the ohmic quench in a metal and also consider photon losses at the boundary of the material. 
By utilizing numerical solutions and Monte Carlo simulations, the presence of condensation has been confirmed, and an energy gap of a few electron volts (eV) per particle has been observed in typical metal crystals with  protons bound to tetrahedral or octahedral sites.
\end{abstract}

\maketitle

\section{Introduction}
The possibility to obtain and manipulate at the nanoscale level coherent and collective quantum optical effects like superradiance and coherent population trapping has been the subject of extensive research over the last two decades \cite{brandes2005coherent,heyl2018dynamical,forn2019ultrastrong,link2020dynamical}.

Superradiance or the Dicke effect occur when an ensemble of molecules or quantum oscillators confined in a sub-wavelength region emit and absorb coherent radiation cooperatively. The coherence among the emitters is mediated by the electric radiation field. In certain systems, the field-matter coupling can be further enhanced by the presence of surface plasmons (plasmonic Dicke effect). A brief review of recent results in this field is given in Sect.\ \ref{plasm}.

In this work we study an idealized system of ${N_{osc}}$ charges oscillating in a lattice, which undergoes a superradiant transition above a certain threshold density. The dynamics of this system involves bulk plasmons. It has been investigated analytically in our previous work \cite{gamberale2023coherent}, where we proved in the large ${N_{osc}}$ limit the existence of an energy gap for a certain set of quantum states, coherent both in the matter and field sectors. Here we perform numerical calculations with a small number of oscillators and we obtain a confirmation of the analytical results, plus some additional insights. In particular, we test the validity of the RWA approximation used in the analytic approach, and we also test the effect of an ohmic quenching term, which is needed if one wants to apply the model to a metal.

In Sect.\ \ref{general} we write the Hamiltonian of the model and we show the main steps leading to the linearization of the electromagnetic interaction through suitable canonical transformations of the photon field operators. This section summarizes some results of \cite{gamberale2023coherent} in a self-consistent way. 

An important characteristic of the model, which makes it suitable for certain applications discussed in \cite{gamberale2023coherent}, is that each oscillator is confined within its electrostatic ``cage'', limiting its oscillation amplitude and causing it to vibrate at the frequency $\omega$, which is the same for all oscillators. In the total Hamiltonian, the frequency $\omega$ is combined with the plasma frequency $\omega_p$ of the oscillating charges to give a dressed frequency $\omega'$, while the photon momentum $\mathbf{k}$ is unchanged. This implies that there exist no states in which the e.m.\ energy generated in the material can propagate to vacuum. 

Another crucial feature is that the coherent e.m.\ modes which arise and oscillate with a fixed phase relation to the matter oscillators have wave vectors pointing in all possible space directions; this enhances the matter-field coupling and makes possible the formation of the energy gap.

In Sect.\ \ref{numerical} we describe the numerical calculation, performed with \texttt{QuTiP} reducing the Hilbert space of the system to a suitable finite vector space and computing the ground state energy $E_0$ either directly, as eigenvalue of $H$, or in the dissipative case via a Monte Carlo simulation. By plotting, as a function of the coupling $\varepsilon=\omega_p/\omega'$, the energy $E_0$ and the expectation and correlation values of the matter and photons fields, a transition to a coherent state is clearly identified.

Finally, Sect.\ \ref{conc} contains our conclusions and outlook.

\section{Plasmonic Dicke effect and superradiance with microcavities and nanoparticles}
\label{plasm}

In this Section we briefly review recent theoretical and experimental results concerning the plasmonic Dicke effect, i.e., coherent superradiance of quantum emitters enhanced by interaction with surface plasmons. This effect occurs in systems confined on a sub-wavelength scale and displays clear analogies to the phenomenon we are going to analyze in this work.

Recent advances in the manipulation of cooperative e.m.\ emission processes at the nanoscale have been summarized by Azzam et al.\ \cite{azzam2020ten} and earlier by Bordo \cite{bordo2017cooperative} with reference to an analytical model of \emph{spaser} (surface plasmon amplification by stimulated emission of radiation), a term originally introduced by Bergman and Stockman \cite{bergman2003surface,stockman2010spaser}. The resonator of a spaser can consist of a metal nanoparticle with size smaller than the wavelength of the involved radiation. The active medium can be for example a semiconductor crystal. 

The surface plasmons of the metal, which in usual applications play the role of concentrating into a small volume the external radiation incident on the metal, in a spaser have the effect of enhancing the coupling among the quantum emitters.

In simpler terms, a spaser requires the confinement of an ensemble of active molecules to a sub-wavelength scale. These molecules emit coherent radiation in a cooperative manner \cite{dicke1954coherence,andreev1980collective,prasad2000polarium,sivasubramanian2001gauge,sivasubramanian2001super,sivasubramanian2003microscopic,prasad2010coherent}), and the emission rates are proportional to the number of molecules.

A related effect occurs for molecules located near a metal nanoparticle; in this case the e.m.\ coupling between molecules is amplified by surface plasmons excited at the nanoparticle surface. 
Pustovit et al.\ \cite{pustovit2009cooperative,pustovit2010plasmon} analyze a system of quantum dipoles close to a metal nanoparticle. They find that the coherent emission of photons by the dipoles is affected by two competing processes: enhancement by resonant energy transfer from excited dipoles to surface plasmons, and quenching by optically inactive excitations in the metal. Simulations indicate that under certain conditions the plasmonic Dicke effect survives non-radiative losses in the metal (see also \cite{delga2014quantum}).

From another point of view, spasers represent an evolution of devices, described in \cite{scheibner2007superradiance}, for coupling quantum dots through their radiation field by embedding them into semiconductor cavities \cite{imamog1999quantum,temnov2005superradiance}; strong coupling between single quantum dots and cavity modes was demonstrated in \cite{reithmaier2004strong,yoshie2004vacuum}. 

Finally, Greenberg and Gauthier \cite{greenberg2012steady} have experimentally demonstrated a collective superradiant instability in a cold atomic vapor pumped by weak optical fields. This results in the emission of multi-mode optical fields in the absence of an optical cavity. The phenomenon is well described by a theoretical model.

\begin{figure}[ht] 
\centering \includegraphics[width=0.8\columnwidth]{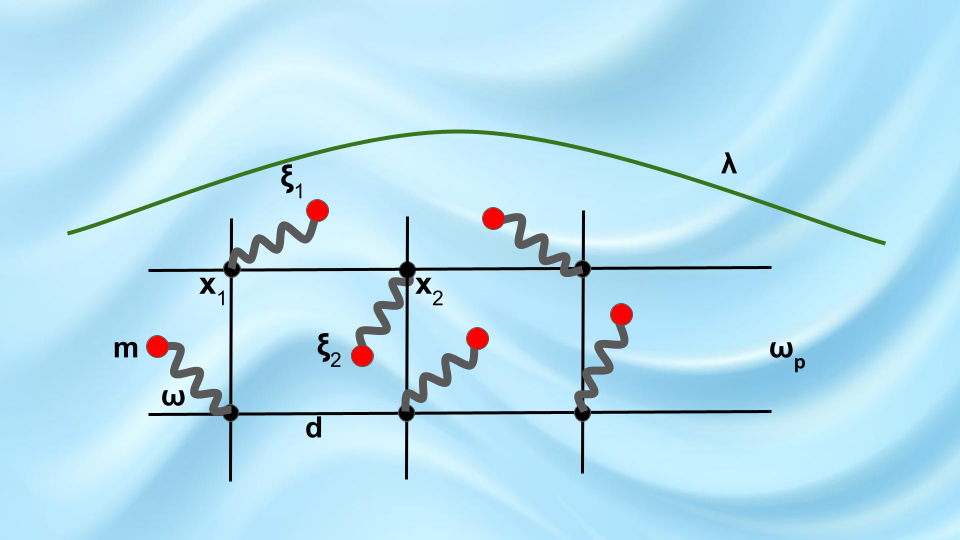}
\caption{
Simplified illustration of the main elements of the model. The crystal lattice has side $d$. At each vertex a charged mass $m$ is bound by an elastic potential with corresponding frequency $\omega$. Coordinates $\mathbf{x}_1$, $\mathbf{x}_2$ etc.\ denote the fixed positions of the vertices. Coordinates $\bm{\xi}_1$, $\bm{\xi}_2$ etc.\ denote the displacements of the masses (their amplitude is exaggerated in the figure). The light blue ripples in the background represent plasma oscillations of the system with frequency $\omega_p$. The effective quadratic Hamiltonian has for each mass $i$ a term proportional to $\frac{1}{2}m\omega^2\bm{\xi}_i^2$ and one proportional to $\frac{1}{2}m\omega_p^2\bm{\xi}_i^2$, which amount to a single oscillator with (dressed) frequency $\omega'=\sqrt{\omega^2+\omega_p^2}$. After inclusion of the e.m.\ field and diagonalization of the total Hamiltonian, it turns out that the e.m.\ modes resonant with the oscillators, one of which is symbolically represented by the green wave, have wavelength $\lambda=2\pi c/\omega$. We suppose $\lambda$ to be such that $d \ll \lambda$.}
\label{illustration}
\end{figure}

\section{The general model}
\label{general}

\subsection{Matter Hamiltonian}

We briefly recall here the model proposed and described in detail in \cite{gamberale2023coherent}. The system we are considering is a cubic lattice hosting charged particles of mass $m$ (see Fig.\ \ref{illustration}),  bound at each vertex by a potential which for small oscillations is harmonic, with elastic coefficient $K_{el}$. The corresponding frequency is denoted by $\omega=\sqrt{K_{el}/m}$. Physical realizations of this idealized system include crystals or metals ``loaded'' with light positive ions, which are typically confined in electrostatic ``cages'' corresponding to tetrahedral holes in the crystal structure \cite{burns1993mineralogical}. 

We consider a finite volume $V$ with $N_{osc}$ positively charged particles with mass $m$ embedded in a neutralizing negative electron density distributed in space (the \emph{jellium crystal}, see Appendix A in \cite{gamberale2023coherent}). Due to the simultaneous presence of the single-particle electrostatic cages and collective plasma oscillations with frequency $\omega_p=\sqrt{\frac{e^2{N_{osc}}}{mV}}$, the Hamiltonian of the $N_{osc}$ charges oscillating about their equilibrium positions can be written in second quantization as 
\begin{equation}
    H_{osc}=\omega' \sum_{n=1}^{N_{osc}} \left[ \bm{a}^\dagger_n(t) \bm{a}_n(t) + \frac{3}{2} \right], \qquad {\rm where} \qquad \omega'=\sqrt{\omega^2+\omega_p^2}
\end{equation}
Here and in the following we use units in which $\hbar=c=1$.

\subsection{Free e.m.\ field}

The Hamiltonian of the quantized e.m.\ field coupled to the system comprises in principle modes of all frequencies, momentum and polarization directions, but the modes which can interact with the oscillators and can therefore be excited are possibly those with frequencies $\omega$, $\omega_p$ or $\omega'$. 

It turns out from the analysis presented in reference \cite{gamberale2023coherent} that, after applying an appropriate canonical transformation to the creation and destruction operators of the electromagnetic field, the modes with $|\mathbf{k}|=\omega$ undergo a frequency renormalization to the value $\omega'$, while simultaneously the $\mathbf{A}^2$-term becomes reabsorbed in the re-definition of the field operators and is no longer present in the resulting Hamiltonian. Consequently, we are left with a simplified Hamiltonian $H_{tot}$ containing only linear terms in the field operators.

Let us start by writing the pure photon part as
\begin{equation}
    H_{phot}=\omega \sum_{p,\mathbf{\hat{k}}} \left[ b^\dagger_{p,\mathbf{k}}(t) b_{p,\mathbf{k}}(t) + \frac{1}{2} \right] \qquad {\rm with} \qquad |\mathbf{k}|=\omega
\end{equation}
In the given expression, the symbol $\sum_{p,\mathbf{\hat{k}}}$ represents the summation over two photon polarizations, denoted by $p$, and the integral over the unit vector $\mathbf{\hat{k}}=\mathbf{k}/\omega$. This integral, in turn, is equivalent to the solid angle component of a three-dimensional integral over wave vectors $\mathbf{k}$. The explicit expression would read as follows:
\begin{equation}
    \sum_{p,\mathbf{\hat{k}}} \to \sum_{p=1}^2 \int d\Omega_\mathbf{\hat{k}}= \sum_{p=1}^2 \int_0^{2\pi} d\phi_\mathbf{\hat{k}} \int_0^{\pi} d\theta_\mathbf{\hat{k}} \sin \theta_\mathbf{\hat{k}}.
\end{equation}

The vector potential can be expressed in the same notation as 
\begin{equation}
    \mathbf{A}(\mathbf{x},t)=\frac{1}{\sqrt{2\omega V}} \sum_{p,\mathbf{\hat{k}}} \left[ b_{p,\mathbf{k}}(t) e^{i\mathbf{k}\mathbf{x}} \bm{ \varepsilon}_{p,\mathbf{k}} + c.c. \right]
    \label{vecpot1}
\end{equation}
where $\bm{ \varepsilon}_{p,\mathbf{k}}$ are polarization vectors. The time dependence takes, in interaction representation, the usual form $b_{p,\mathbf{k}}(t)=b_{p,\mathbf{k}}e^{-i\omega t}$. For simplicity we will omit the time dependence of the operators in the following.

We assume that the finite volume $V$ we are considering is such that $V \ll \lambda^3$, where $\lambda=\frac{2\pi}{\omega}$. In concrete physical systems, the frequency $\omega$ is typically on the order of the largest frequency of optical phonons, around 0.1 eV. It follows that the parameter ${N_{osc}}$ can be quite large, reaching values on the order of $10^9$ or even higher. In \cite{gamberale2023coherent} we presented analytical variational calculations applicable in the regime of large ${N_{osc}}$. However, in the current study, we will focus on providing numerical solutions for small values of ${N_{osc}}$.

\subsection{Matter-field interaction}

The field-matter interaction terms are obtained as usual through the gauge-invariant extension of the momentum $\bm{p} \to \bm{p}+e\mathbf{A}$ in the matter Hamiltonian. This gives two terms, which in \cite{gamberale2023coherent} are written together, while here we list them separately. The dipole interaction term has the form
\begin{equation}
    H_{dip}=\sum_{n=1}^{N_{osc}} \frac{e}{m} \bm{p}_n \mathbf{A}(\mathbf{x}_n+\bm{\xi}_n,t)
\end{equation}
where $\bm{p}_n$ is the momentum of the $n$-th oscillator (expressible through the operators $\bm{a}_n$, $\bm{a}^\dagger_n$). Furthermore, there is a ``diamagnetic'' term with the square of the vector potential
\begin{equation}
    H_{\mathbf{A}^2}=\sum_{n=1}^{N_{osc}} \frac{e^2}{2m} \mathbf{A}^2(\mathbf{x}_n+\bm{\xi}_n,t)
\end{equation}
The total Hamiltonian is
\begin{equation}
    H_{tot}=H_{osc}+H_{phot}+H_{dip}+H_{\mathbf{A}^2}
\end{equation}
In the argument of $\mathbf{A}$, $\mathbf{x}_n$ denotes the equilibrium position of the $n$-th oscillator, and $\bm{\xi}_n$ denotes the displacement with respect to this equilibrium position. In the following, however, the space dependence of $\mathbf{A}$ will be neglected (``dipole approximation''), since we assume that the size $V^{1/3}$ of the system is much smaller than $\lambda$.

With a linear transformation of the photon operators of the form 
\begin{equation}
    c_{p,\mathbf{k}}=\frac{1}{2\sqrt{\omega \omega'}} \left[ (\omega'+\omega)b_{p,\mathbf{k}}+(\omega'-\omega)b^\dagger_{p,\mathbf{k}}  \right]
    \label{trasf-b-c}
\end{equation}
the total Hamiltonian can be simply rewritten in terms of the dressed operators $c$ and $c^\dagger$ as
\begin{equation}
    H_{tot}=H_{osc}+\omega'\sum_{p,\mathbf{\hat{k}}} \left( c^\dagger_{p,\mathbf{k}} c_{p,\mathbf{k}} + \frac{1}{2} \right) + \frac{i\omega_p}{2\sqrt{{N_{osc}}}} \sum_{n=1}^{N_{osc}} \sum_{p,\mathbf{\hat{k}}} \left[ 
\bm{a}^\dagger_n-\bm{a}_n \right] \left[ c_{p,\mathbf{k}} \bm{\varepsilon}_{p,\mathbf{k}} + c.c. \right]
\label{h9}
\end{equation}

The vector potential can be expressed, in terms of the new operators, as
\begin{equation}
    \mathbf{A}(\mathbf{x},t)=\frac{1}{\sqrt{2\omega' V}} \sum_{p,\mathbf{\hat{k}}} \left[ c_{p,\mathbf{k}} e^{i\mathbf{k}\mathbf{x}} \bm{ \varepsilon}_{p,\mathbf{k}} + c.c. \right]
\end{equation}
which is the strict analogue of (\ref{vecpot1}); however, it is interesting to note that the dispersion relation of the dressed photons created and destroyed by $c^\dagger$ and $c$ is $\omega'=\sqrt{k^2+\omega^2}$, different from the vacuum dispersion relation $k=\omega$.

It is clear from (\ref{trasf-b-c}) that the transformation from the operators $b$, $b^\dagger$ to $c$, $c^\dagger$ is well defined only if $\omega \neq 0$. Therefore the harmonic potential which bounds all charges to the lattice sites with the same $\omega$ is an essential element of the model. A similar operator transformation is employed in cavity QED \cite{faisal1987theory}. In that context, $\omega$ is defined by the optical cavity that contains the system. For linear cavities, the direction of $\mathbf{k}$ is also fixed. Here, there are many directions of $\mathbf{k}$ contributing to the interaction and in the next subsection we will apply a geometrical projection operation plus a summation over $\mathbf{\hat{k}}$ to simplify the algebraic manipulation of the photon operators.

\subsection{Projected and $k$-summed photon operators}

It is convenient to rewrite the Hamiltonian using operators of creation and destruction of the photons-plasmons which are projected along the 3 space directions defined by the unit vectors $\mathbf{\hat{e}}_i$ ($i=1,2,3$) and summed over the possible directions $\mathbf{\hat{k}}$ of momentum.

We shall call these operators $C^\dagger_i$ and $C_i$ ($i=1,2,3$); they are defined by linear combinations of the operators $c_{p,\mathbf{k}}$ with coefficients given by scalar products between the unit vectors $\mathbf{\hat{e}}_i$ and the polarization vectors $\bm{ \varepsilon}_{p,\mathbf{k}}$:
\begin{equation}
    C_i=\sqrt{\frac{3}{8\pi}} \sum_{p,\mathbf{\hat{k}}} ( \mathbf{\hat{e}}_i  \bm{ \varepsilon}_{p,\mathbf{k}} ) c_{p,\mathbf{k}}.
\end{equation}
Thanks to the normalization factor $\sqrt{\frac{3}{8\pi}}$, the states created and destroyed by these operators turn out to be correctly normalized, and the operators themselves satisfy the commutation relations
\begin{equation}
    [C_i,C^\dagger_j]=\delta_{ij}; \qquad [C_i,C_j]=[C^\dagger_i,C^\dagger_j]=0
\end{equation}

Using the operators $C_i$, $C^\dagger_i$ we can rewrite the vector potential  and the Hamiltonian (\ref{h9}) (in dipole approximation) as follows:
\begin{equation}
    \mathbf{A}=\frac{1}{\sqrt{2\omega' V}} \sum_{i=1}^3 (C_i+C^\dagger_i) \mathbf{\hat{e}}_i
\end{equation}
\begin{equation}
    H_{tot}=H_{osc}+\omega'\sum_{i=1}^3 \left( C^\dagger_i C_i + \frac{1}{2} \right) + \frac{i\omega_p}{2\sqrt{{N_{osc}}}} \sqrt{\frac{3}{8\pi}} \sum_{n=1}^{N_{osc}} \sum_{i=1}^3 \left[ a^\dagger_{n,i}C_i - a_{n,i}C^\dagger_i 
    + a^\dagger_{n,i}C^\dagger_i - a_{n,i}C_i\right] 
\label{H-finale}
\end{equation}
where $a_{n,i}$ denotes the $i$-th spatial component of $\bm{a}_n$.

\subsection{Summary of analytical results}
\label{summary}

In our work \cite{gamberale2023coherent} we have studied analytically the properties of the Hamiltonian (\ref{H-finale}) in the RWA approximation, i.e., disregarding the terms of the form $aC$ or $a^\dagger C^\dagger$, and in the limit of large ${N_{osc}}$. In particular, we have shown that a threshold value of the coupling $\varepsilon=\omega_p/\omega'$ exists, above which the ground state of the system is not the usual perturbative ground state where all oscillators have excitation number zero. For this purpose we have defined a set of trial quantum states that take the form of coherent states, both in the degrees of freedom of the charged oscillators and in those of the field oscillator. 

The minimum energy is achieved when the charged oscillators all have the same phase, and the field oscillator a phase which differ by $\pi/2$. More precisely, above the threshold value of $\varepsilon$ the Hamiltonian is a quadratic form with negative modes proportional to $|\alpha|^2$ (the squared amplitude of the coherent trial states of the charges), so that the system becomes unstable when $|\alpha|>0$, as we shall see later in more detail.

Following a heuristic physical argument we imposed a limiting value $\alpha_{max}$ to the coherent oscillation amplitude $\alpha$, by assuming that each charge is confined in one of the lattice cells. In this way it is possible to compute the energy per particle $E_0$ of the ground state as a function of $\varepsilon$, up to second order in the Brillouin-Wigner approximation. The resulting expression is 
\begin{subequations}
 \begin{empheq}[left={\empheqlbrace\,}]{align}
    E_0^{(1)} =& \omega'
|\alpha_{max}|^2
\left(1-\frac{2\pi}{3}
\varepsilon^2
\right)\ \ \ \text{first order}
\label{eq:energy-gap1}
\\
    E_0^{(2)} =& \omega'
|\alpha_{max}|^2
\left(1-\frac{8\pi}{3}
\varepsilon^2
\right)\ \ \ \text{second order.}
\label{eq:energy-gap2}
\end{empheq}
\label{eq:energy-gap}
\end{subequations}
The corresponding critical coupling is $\varepsilon_{crit}^{(1)}=\sqrt{\frac{3}{2\pi}}\simeq 0.69$ to first order perturbation theory and $\varepsilon_{crit}^{(2)}=\sqrt{\frac{3}{8\pi}}\simeq 0.35$ to second order.

Considering the case of protons located in the lattice cells of a metal, by replacing the appropriate values in Eq. \eqref{eq:energy-gap}, the energy gap per particle $|E_0|$ turns out to be of the order of a few eV, much larger than the average thermal excitation energy, implying that such states are thermally stable. We recall that the plasma frequency $\omega_p$ depends on the density of the oscillating charges and the bare oscillation frequency $\omega$ depends of the electrostatic forces in the crystal; both are of the order of $10^{13}$ Hz.
In Appendix A of \cite{gamberale2023coherent} a model of the electrostatic force called \emph{jellium crystal} has been developed where the frequencies, $\omega$, $\omega_p$ and $\omega'$ have been computed.
In the case of a metal with lattice spacing $d=2.5\angstrom$ loaded with protons in each lattice site we find $\omega=0.35$ eV, $\omega_p=0.22$ eV, $\omega'=0.41$ eV, $\varepsilon=0.52$.

\section{Numerical analysis}
\label{numerical}

In order to reach a better understanding of the model and to check numerically its validity in a finite-dimensional case, we have implemented a reduced version in \texttt{QuTiP}, the Python toolbox widely used for quantum mechanics and especially for quantum optics \cite{qutip1,qutip2}.

\subsection{Setting up the Hamiltonian with \texttt{QuTiP}}

We have written a second-quantized unperturbed Hamiltonian $H_0$ which describes one harmonic oscillator with destruction/creation operators \texttt{C}, \texttt{C.dag()} representing one of the photon modes of (\ref{H-finale}), plus a small number $N_{osc}$ of ``material'' oscillators with destructors/creators \texttt{a1}, \texttt{a2}, ... \texttt{a1.dag()}, \texttt{a2.dag()}, ... having the same frequency of the photons. The energy calculations are in units of $\omega'$. We then added a coupling term with a coefficient given by the 3D enhancement factor $\sqrt{\frac{8\pi}{3}}$, as already discussed, and by the coupling constant $0<\varepsilon<1$ whose value defines below-threshold and above-threshold conditions for the condensation to the coherent state. 

After defining the total Hamiltonian, one can directly obtain with \texttt{QuTiP} the energy of the ground state (and also that of the excited states) as a function of the coupling $\varepsilon$. However, there is also a dependence on the dimension of the finite vector space which approximates the full Hilbert space in describing the states of the various oscillators. We denote with $N_{exc}$ the highest excitation level of each of the particle oscillators, and with $N_{phot}$ the highest level of the photon oscillator. They can be varied independently, but physically it is reasonable to assume first that $N_{phot}$ is equal to $N_{exc} \cdot {N_{osc}}$,  since we expect that each energy level of each oscillator emits and absorbs a photon.

% Through the analysis of the ground state energy calculation illustrated in Fig.\ \ref{fig:stabiliz}, which depicts the energy gap as a function of $N_{exc}$, it becomes apparent that when $\varepsilon$  is adjusted to a value greater than the threshold, the negative energy gap with respect to the unperturbed state of the oscillators grows unrestricted as $N_{exc}$ progressively increases.

\begin{figure}[ht] 
\centering \includegraphics[width=0.8\columnwidth]{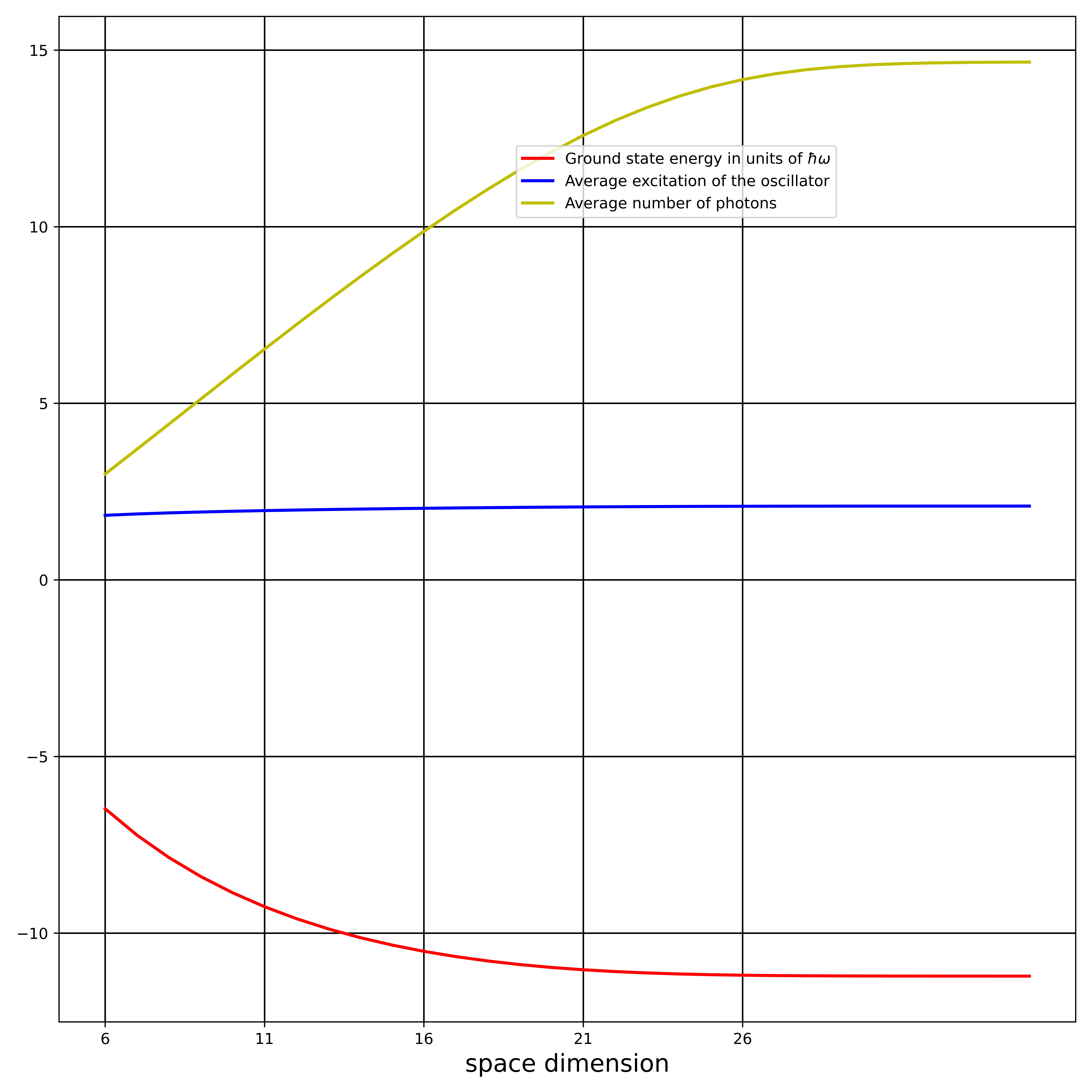}

\caption{Dependence on the vector space dimension $N_{exc}$ with $\varepsilon=1$ of (1) the average excitation number of one harmonic oscillator confined with a term $(a^\dagger)^4 a^4$; (2) the average excitation number of a photon oscillator coupled with it; (3) the ground state energy of the system. Note the convergence to finite values when $N_{exc}$ is approximately greater than 35.}
\label{fig:stabiliz}
\end{figure}

By examining the computation of the ground state energy shown in Fig. \ref{fig:stabiliz}, which displays the energy gap as a function of $N_{exc}$, it becomes evident that as $\varepsilon$ is tuned to a value higher than the threshold $\varepsilon_{crit}=0.35$, the negative energy gap grows without limitations as $N_{exc}$ increases.

When the phase transition occurs,the average occupation number of all oscillators (particles and photons) reaches a level of approximately 50\% of the maximum possible excitation fixed by the chosen finite space dimension $N_{exc}$. This clearly shows that all accessible levels are occupied, while for a correct description of the system with a finite dimension of the Hilbert space of the states one usually requires that the upper levels are essentially unexcited, so that physical quantities are independent from $N_{exc}$. 

This lack of a lower bound for the energy is consistent with the analytical model \cite{gamberale2023coherent}, where the Hamiltonian in conditions above threshold allows for coherent trial states that depend on a "classical" amplitude $\alpha$ of the particle oscillators. In that case, $\alpha$ is not bounded, and the energy gap is proportional to $|\alpha|^2$ (see Sect.\ \ref{sec:spacedim}).
The correlation between the numerical outcome and the selected dimension of the vector space raises concerns regarding the accuracy of the calculation. In order to tackle this problem, we will introduce an additional component into the Hamiltonian. This supplementary component has a minimal impact on the energy per particle when $\alpha$ is small, but it effectively limits the extent of charge oscillations. By employing this approach, higher modes of excitation will remain unexcited, leading to computational results that become unaffected by the selected dimensionality once a critical threshold is exceeded.
This matter will be discussed further in subsection \ref{sec:spacedim}.

We stress that the coupling constant $\varepsilon$ arises from the existence of a significant number of high-density oscillators within the system. In the scenario where the physical system consists of only a small amount of oscillating charges distributed in the volume $V$, the coupling with the electromagnetic field would be minute, incapable of inducing any form of phase transition. In the present simulation, due to the limited computation power, we examine the behavior of a \emph{restricted} number of oscillators, investigating their interaction with the electromagnetic field and the corresponding state of minimum energy. The presence of the large number of oscillators is incorporated in the parameter $\varepsilon>0$ which, depending on the plasma frequency $\omega_p$, reflects a high density of charged oscillators.

\begin{figure}[ht] 
\centering \includegraphics[width=0.8\columnwidth]{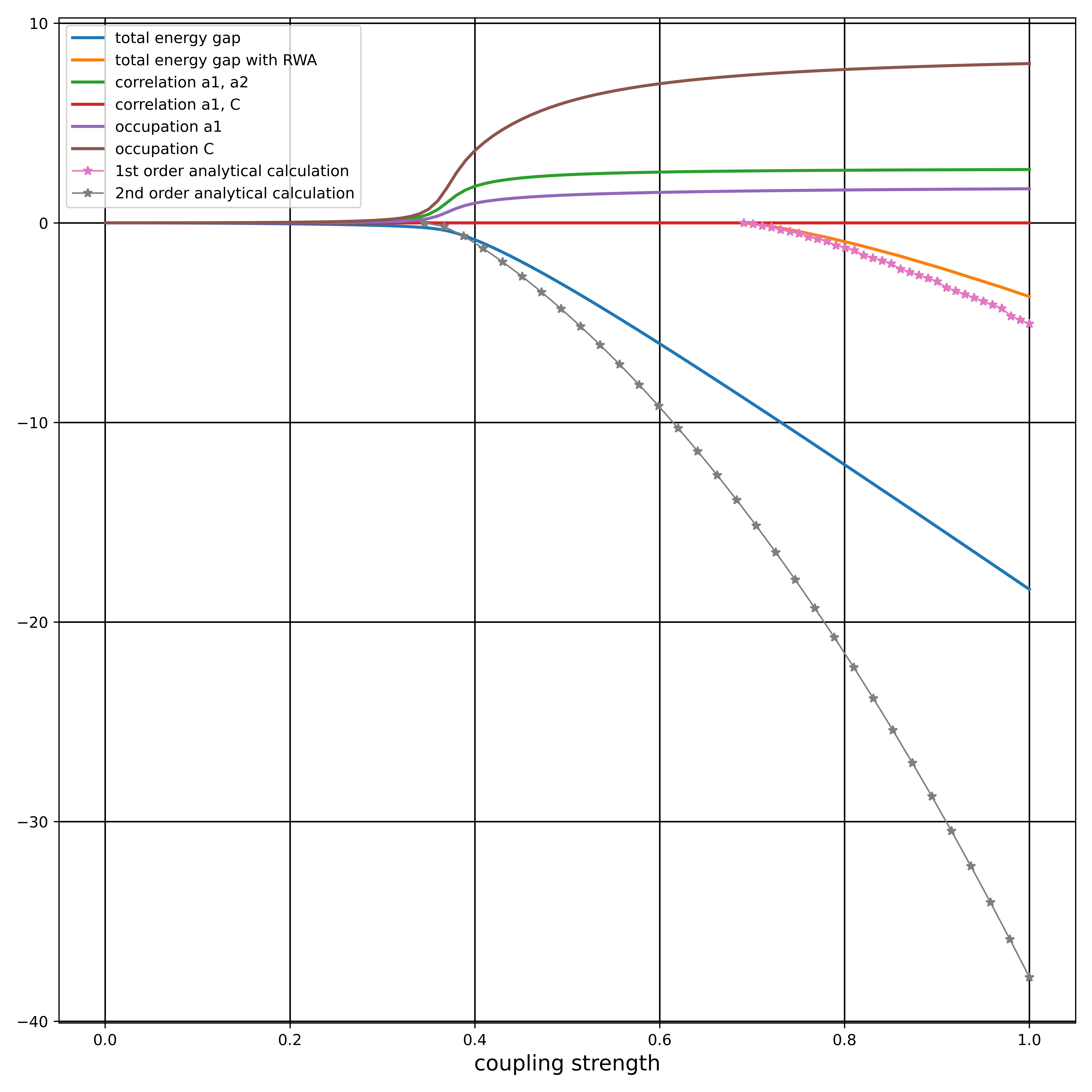}
\caption{
Dependence on the adimensional coupling $\varepsilon$ of the ground state energy (``energy gap'') and of various correlations and occupation numbers, showing the occurrence of condensation to a coherent state at $\varepsilon \simeq 0.35$. The system comprises 3 material oscillators ($N_{osc}=3$). The dimensions of the vector spaces used for the numerical solutions are such to meet the stabilization condition (compare Fig.\ \ref{fig:stabiliz}). The energy gap in the RWA approximation is also displayed for comparison. The correlations and occupation numbers shown can be explicitly written respectively as $\langle a_1^\dagger a_2+a_1 a_2^\dagger \rangle_0$ (correlation between two material oscillators), $\langle a_1^\dagger C+a_1 C^\dagger \rangle_0$ (correlation between one material oscillator and the photon oscillator), $\langle a_1^\dagger a_1 \rangle_0$
(occupation number of one material oscillator) and $\langle C^\dagger C \rangle_0$ (occupation number of the photon oscillator).}
\label{fig:gap-et-al}
\end{figure}

In Fig. \ref{fig:gap-et-al} is represented the result of the numerical calculation of the ground state energy for various values of $\varepsilon$ together with the average occupation numbers of the photon and matter field.
The numerical solutions have been obtained both with and without the RWA approximation and show that the transition to the coherent state is present in both cases, although with different threshold values for the coupling $\varepsilon$.
More specifically, below threshold the energy of the ground state is very close to zero, while above threshold it becomes negative and proportional to the number of material oscillators (see Table \ref{table1}). 

\begin{figure}[ht] 
\centering \includegraphics[width=0.8\columnwidth]{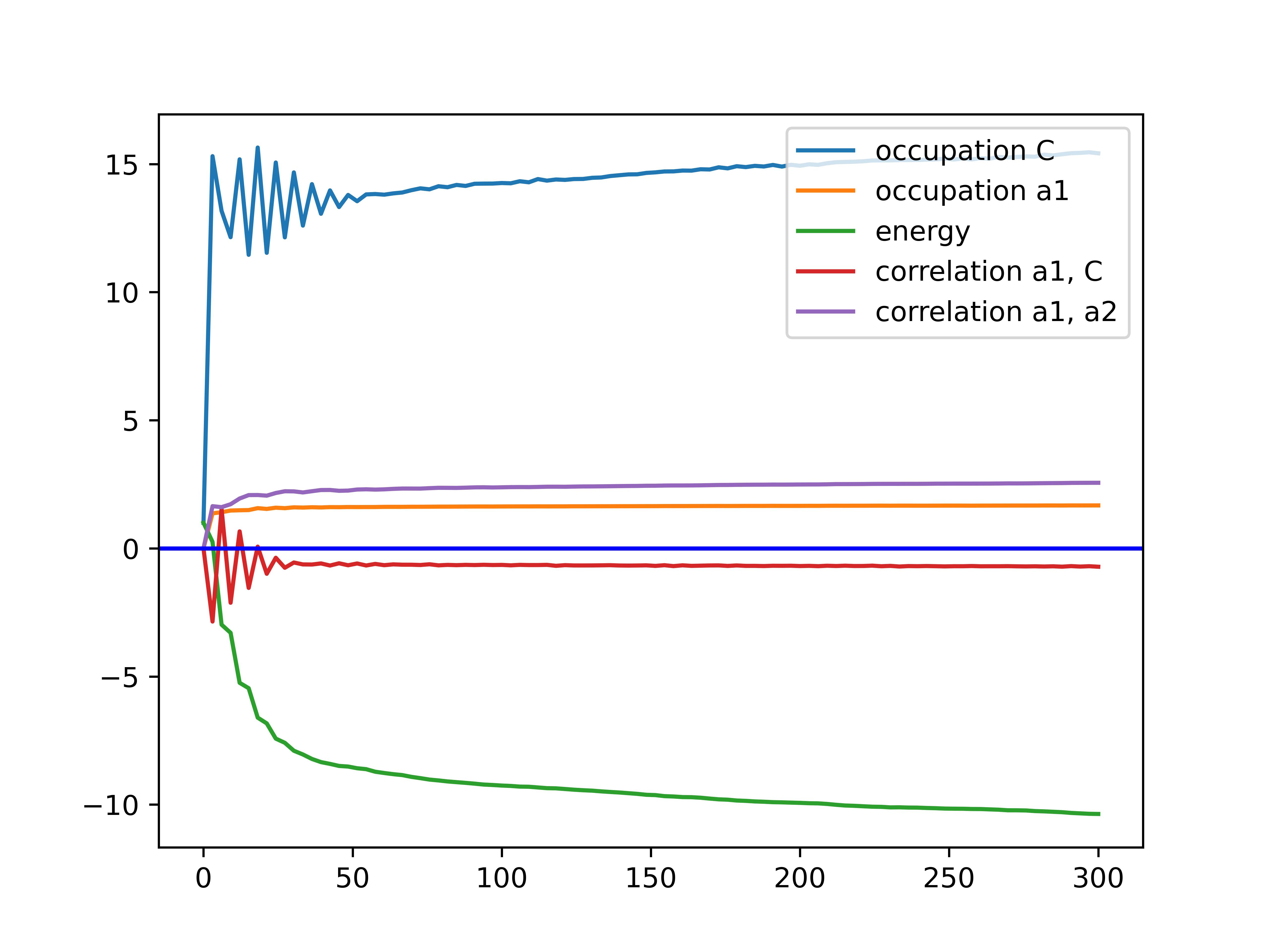}
\caption{Result of a Monte Carlo simulation with dissipative term. Three material oscillators, $N_{phot}=60$. Dissipation operator \texttt{0.4*C}. See the caption of Fig.\ \ref{fig:gap-et-al} for the full expressions of the correlations and occupation numbers. Note the negative correlation between the material oscillator and the photon oscillator.}
\label{fig:monte-carlo-dissip}
\end{figure}

\begin{figure}[ht] 
\centering \includegraphics[width=0.8\columnwidth]{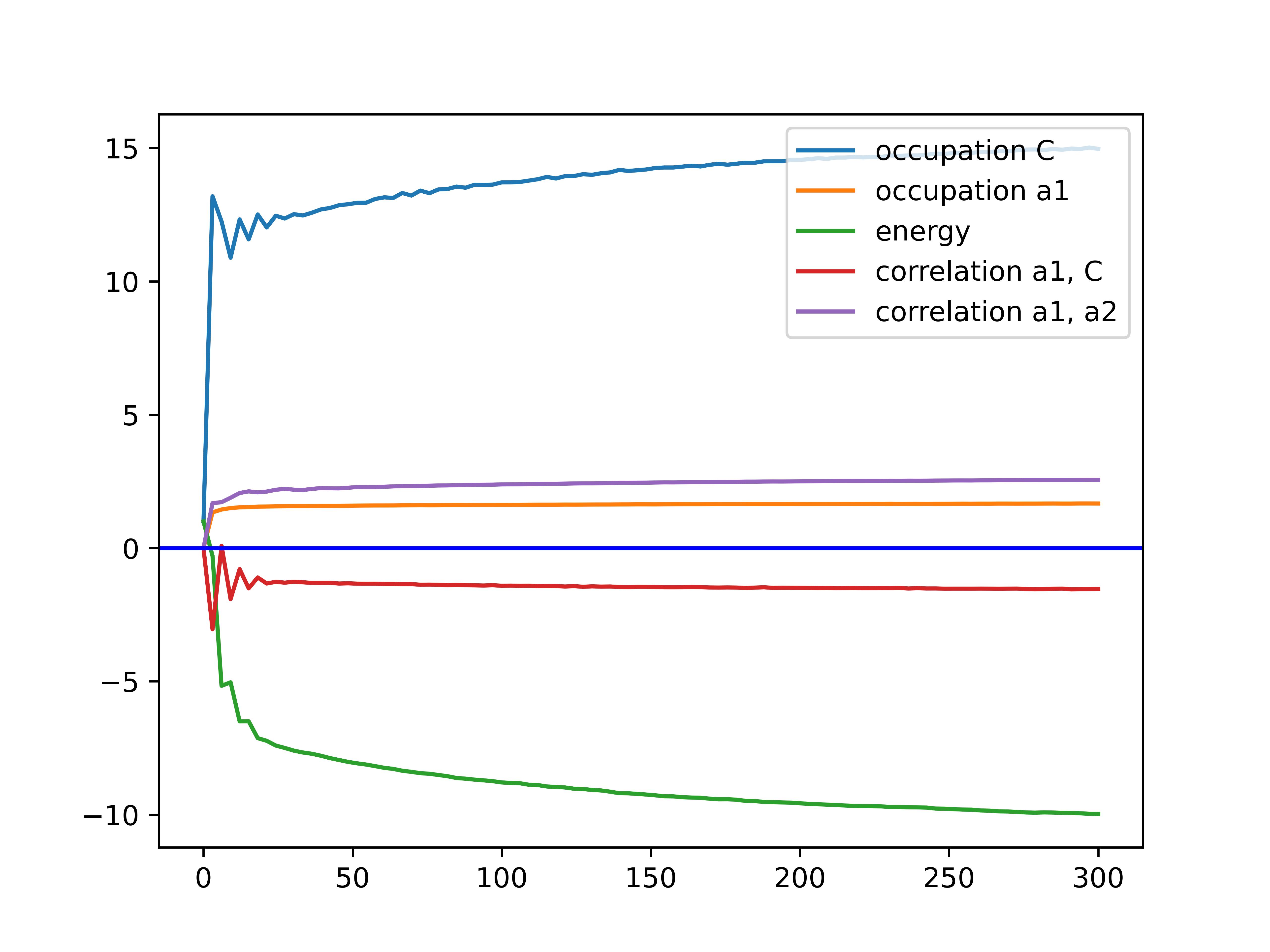}
\caption{Result of a Monte Carlo simulation with greater dissipation than in  Fig.\ \ref{fig:monte-carlo-dissip}. Dissipation operator \texttt{0.6*C}. Note that the negative correlation between material oscillator and photon oscillator increases, in absolute value, when dissipation increases, while the ground state energy remains the same, asymptotically $\simeq -11$, as from the spectrum without dissipation. 
}
\label{fig:monte-carlo-dissip2}
\end{figure}

It is intriguing to observe that, while in the case of the RWA the system's energy is rigorously zero below the threshold $\varepsilon_{crit}=0.69$, as analytically calculated in \cite{gamberale2023coherent} (Fig. \ref{fig:gap-et-al}, orange line), in the absence of the RWA approximation, the energy is slightly negative even for values of $\varepsilon<\varepsilon_{crit}$ (Fig. \ref{fig:gap-et-al}, blue line). This phenomenon has been discussed and termed as \emph{weak coherence} in \cite{prep-weak-coherence}.

As a final note, we observe that the diagonalization of the Hamiltonian in the RWA in \texttt{QuTiP} shows a threshold for symmetry breaking at $\varepsilon_{crit}=\varepsilon_{crit}^{(1)}$ instead of the more accurate value $\varepsilon_{crit}^{(2)}$.
The reason behind this phenomenon can be attributed to the conservation of the total number of photons and oscillator excitation levels in the interaction term within the RWA. However, the full interaction term does not exhibit this conservation. As a consequence, the second-order contribution within the RWA accurately captures the exchange of photons between \emph{different} oscillators. Considering that the calculation involves a very small number of oscillators, the second-order term becomes insignificant. However, this argument does not hold true for the full interaction term, as it mixes states where the sum of the number of photons and excitation levels remains constant. Consequently, the second-order term becomes significant even with a few oscillators, as observed in Fig. \ref{fig:gap-et-al}.

\subsection{Limiting the vector space dimension with oscillator ``cages'' 
\label{sec:spacedim}}

A physically meaningful model requires the implementation in the total Hamiltonian of a reasonable limitation mechanism for the amplitude of the oscillators. In \cite{gamberale2023coherent} we introduced for this purpose electrostatic ``cages'' which set an upper limit on the coherent oscillation amplitude $\alpha$. It is therefore supposed that the oscillating charges bound on each lattice site cannot move to another cell of the lattice. A possible way of implementing this condition here numerically is to add in the Hamiltonian for each harmonic oscillator a term of the form $(a^\dagger)^n a^n$, with $n$ sufficiently large (e.g., $n=4$). This term acts in practice as a high potential barrier, leaving the lowest energy levels of the oscillator unchanged but causing a large increase in the energy of the states with large excitation number. One can check, for example, that with $n=4$ the energy levels are (apart from the offset $\frac{1}{2}$) equal to $(0,1,2,3,28,125,366,...)$, while the corresponding expectation values of the operator $\hat{x}^2$ are the same of a pure harmonic oscillator, namely $(\frac{1}{2},\frac{3}{2},\frac{5}{2},\frac{7}{2},\frac{9}{2},...)$, implying a strong spatial confinement.

With this modification, $H$ is bounded from below and the excitation level of the harmonic oscillators converges to a finite value as the size of the base used increases. It is important to note that when the amplitude of material oscillators is limited by a term of the type $(a^\dagger)^n a^n$, the amplitude of photon oscillators becomes also limited, without the need for a corresponding term. See example in Fig.\ \ref{fig:stabiliz}.

An alternative approach for limiting the amplitude of oscillations involves considering that the binding potential wells, responsible for confining each material particle within the lattice, possess finite depth instead of being infinitely deep. Beyond the bound states, there exists a continuous spectrum characteristic of free particles or Bloch quantum states. It is evident from a physical standpoint that excitations towards these free states do not contribute to coherence as their frequencies are not in resonance with the photon field. Hence, the states relevant for superradiance are, at least in terms of order of magnitude, roughly equal in number to the bound states. However, they are actually fewer due to the non-uniform energy spacing of states in close proximity to the continuum spectrum. The precise estimation of the number of states involved in the coherence process can be accomplished with reasonable accuracy, but we defer this calculation to a future study.

Table \ref{table1} presents an illustrative set of computed ground state energy values, showcasing the dependence on the number of material oscillators, with a fixed coupling strength of $\varepsilon=0.7$. The data were obtained with the method \texttt{H.groundstate}, where for ex.\ for $N_{osc}=2$ we have \texttt{H=H0+Hint+H8}, with 

\noindent
\texttt{H0 = a1.dag()*a1 + a2.dag()*a2 + C.dag()*C}, 

\noindent
\texttt{Hint = sqrt(8*np.pi/3)/sqrt(Nosc)*(1j/2)*(C + C.dag())*(a1.dag()-a1+a2.dag()- a2)}, 

\noindent
\texttt{H8 = a1.dag()**4 * a1**4 + a2.dag()**4 * a2**4}.

The second row of Table \ref{table1} displays the chosen space dimension of the photon field $N_{phot}$ used in the numerical calculation, according to the ``stabilization'' criterion illustrated in Fig.\ \ref{fig:stabiliz}, but with a stabilization value $N_{phot}=20\cdot N_{osc}$, suitable for coupling $\varepsilon=0.7$. The value of the energy for $N_{osc}=5$ has been computed with a Monte Carlo algorithm including a dissipation term, as described in Sect.\ \ref{MC} (see also Figs.\ \ref{fig:monte-carlo-dissip}, \ref{fig:monte-carlo-dissip2}), since the multiple tensor products needed for the description of the state lead to a dimension of the Hamiltonian matrix which would require to allocate 156 GiB for an array with shape (102400,102400) and data type \texttt{complex128}. 

In the numerical solution of certain coherent systems, symmetry properties allow to reduce the dimension of the vector space; see e.g.\ \cite{shammah2018open}. In our case, this does not appear to be possible, and the $N_{osc}$ harmonic oscillators must be left free to evolve independently, like in other calculations concerning quantum synchronization (\cite{tindall2020quantum} and refs.).

\begin{table}
\begin{center}
\begin{tabular}{|c|c|c|c|c|c|} 
\toprule
$N_{osc}$ & 1 & 2 & 3 & 4 & 5 \nonumber \\
\hline
$N_{phot}$ & 20 & 40 & 60 & 80 & 100 \nonumber \\
$E_0$ & -3.78 & -7.56 & -11.3 & -15.1 & -18 $\pm 1$ \nonumber \\
\hline
\end{tabular}	
\caption{Ground state energy $E_0$ in $\hbar \omega$ units (third row) in dependence on the number of material oscillators $N_{osc}$, for coupling strength $\varepsilon=0.7$. The second row gives the maximum excitation level of the photon field used in the numerical calculation. The value of the energy for $N_{osc}=5$ has been computed with a Monte Carlo algorithm including a dissipation term.
}		
\label{table1}
\end{center}
\end{table}

\subsection{Monte Carlo simulations in the dissipative case}
\label{MC}

As mentioned in Sect.\ \ref{plasm} in relation to the plasmonic Dicke effect, it is generally important to check how the ohmic quench influences the onset of superradiance in a system, or the condensation of the system to a coherent state. In our case, this check allows us to apply the model not only to a dielectric crystal, but also to a metal. Moreover, dissipation can occur at the external boundary of the active material, because the dressed photon frequency $\omega'$ is greater than the frequency of photons with the same momentum in vacuum, and therefore some photons can escape the material if they can release the excess energy at the surface.
Following \cite{pustovit2009cooperative}, a dissipation term is added with the form
\begin{equation}
    L_{dis}= \omega'\Gamma C
\end{equation}
where $\Gamma$ is the inverse of the decay constant in units of $\omega'$.
A numerical method implemented in \texttt{QuTiP} for simulating the dissipative evolution of a quantum system is the Lindblad master equation. However, when the dimension of the vector space of the states grows, the Lindblad solver called \texttt{mesolve} tends to fail due to a loss of convergence or to insufficient memory. 

An alternative to \texttt{mesolve} is the Monte Carlo solver \texttt{mcsolve}, which exhibits more favorable scalability with dimensionality. However, it comes at the cost of introducing statistical errors and requiring long simulation runs. Figures \ref{fig:monte-carlo-dissip} and \ref{fig:monte-carlo-dissip2}  illustrate the outcomes of Monte Carlo simulations for the scenario involving three charged oscillators ($N_{osc}=3$) in addition to the photon oscillator. Two different dissipation levels are specified in the captions, corresponding to above-threshold conditions ($\varepsilon=0.7$). The presence of coherent condensation is indicated by a negative energy gap, which, within the margin of error, matches the gap computed in Fig.\ \ref{fig:gap-et-al} using the method \texttt{.groundstate()}.
Notably, this outcome remains independent of the dissipation level. The interpretation is as follows: the dissipation term introduces an imaginary component, and once symmetry breaking occurs, it is counterbalanced by an opposite-signed imaginary component originating from the interaction term. This interaction term incorporates a phase difference between the matter and field, deviating from the analytically calculated phase shift of $\frac{\pi}{2}$ between the oscillators and the field by an amount
\begin{equation}
\sqrt{\frac{8\pi}{3}}\epsilon |\mathcal{A}\alpha_1|\sin\theta = -\frac{1}{2}\langle a_1^\dagger C + a_1C^\dagger \rangle_0,
\label{eq:dissip1}
\end{equation}
where $\frac{\pi}{2} + \theta$ represents the phase difference between the field and the oscillator. The term $\frac{1}{2}\langle a_1^\dagger C + a_1C^\dagger \rangle_0$ denotes the matter-field correlation, which is zero in the absence of dissipation due to the $\pi/2$ phase difference between the matter and field oscillators (Fig. \ref{fig:gap-et-al}, red line). However, it becomes negative in the presence of dissipation (Fig. \ref{fig:monte-carlo-dissip}, red line) and becomes more pronounced with increasing $\Gamma$ (Fig. \ref{fig:monte-carlo-dissip2}, red line).

Also essentially independent from the dissipation level are the average occupation numbers of the oscillators and the correlations between charged oscillators like $\langle a_1^\dagger a_2+a_1 a_2^\dagger \rangle_0$. 

\section{Conclusions}
\label{conc}
In conclusion, we have successfully demonstrated, through a numerical calculation, the occurrence of a phase transition in a system comprising positively charged oscillators immersed in a neutralizing negatively charged medium and interacting with the electromagnetic field. This phase transition occurs when a density threshold of the oscillators is reached. Our previous work \cite{gamberale2023coherent} analytically identified this phase transition using the RWA. The current study not only confirms the validity of the analytical result but also extends its applicability by considering the case without RWA.

Furthermore, we have investigated the impact of the finite dimension of the vector space employed in the numerical calculation. We found that this finite dimension has no bearing on the manifestation of the phase transition. To ensure the boundedness of the matter field's amplitude throughout the numerical simulation, we have introduced a positive definite term into the Hamiltonian. This additional term has a negligible impact on the occurrence of the phase transition itself, while simultaneously constraining the maximum oscillation amplitude to a value that remains independent of the dimensionality of the vector space.

It is noteworthy that a crystal characterized by positively charged ions bound to lattice sites and oscillating harmonically at the same frequency exhibits captivating resemblances to systems commonly encountered in cavity quantum electrodynamics (CQED). The emergence of long-range coupling among the ions stems from their interaction with a photon field, where the energy-momentum relationship of the photons is modified due to the presence of plasma oscillations. Through the utilization of a canonical operator transformation, the system's Hamiltonian can be diagonalized, facilitating the identification of conditions that give rise to the phenomenon of superradiance. Unlike systems typically associated with the Dicke effect, such as ensembles of spins or two-level molecules, this crystal system more closely resembles a set of quantum oscillators, with their states spanning a larger vector space in general.

Furthermore, we have demonstrated that condensation to the coherent state persists even in the presence of strong dissipation. Interestingly, we find that the energy gap between the coherent ground state and the perturbative ground state of the oscillators plus field remains unaffected by dissipation. However, the correlation between the charged oscillators and the photon oscillator does exhibit a dependence on the level of dissipation.

Future investigations will address the following key aspects: (1) A more realistic characterization of the potential wells that confine the ions to lattice sites, offering a more comprehensive understanding of the system. (2) An in-depth analysis of the coherence properties of the numerically obtained ground state in comparison to the coherent trial states employed in previous analytical calculations, which utilized the RWA. While the energy gap in both cases is of similar magnitude, the coherence properties of the trial states appear to be stronger. (3) Investigation of the excited states of the system and their associated coherence properties, providing an understanding of the system's behavior beyond the ground state. (4) A scattering theory of the coherent states.

\bibliography{numerical_analysis} 

\begin{thebibliography}{10}

\bibitem{brandes2005coherent}
T.~Brandes, ``Coherent and collective quantum optical effects in mesoscopic
  systems,'' {\em Physics Reports}, vol.~408, no.~5-6, pp.~315--474, 2005.

\bibitem{heyl2018dynamical}
M.~Heyl, ``Dynamical quantum phase transitions: a review,'' {\em Reports on
  Progress in Physics}, vol.~81, no.~5, p.~054001, 2018.

\bibitem{forn2019ultrastrong}
P.~Forn-D{\'\i}az, L.~Lamata, E.~Rico, J.~Kono, and E.~Solano, ``Ultrastrong
  coupling regimes of light-matter interaction,'' {\em Reviews of Modern
  Physics}, vol.~91, no.~2, p.~025005, 2019.

\bibitem{link2020dynamical}
V.~Link and W.~T. Strunz, ``Dynamical phase transitions in dissipative quantum
  dynamics with quantum optical realization,'' {\em Physical Review Letters},
  vol.~125, no.~14, p.~143602, 2020.

\bibitem{gamberale2023coherent}
L.~Gamberale and G.~Modanese, ``Coherent plasma in a lattice,'' {\em Symmetry},
  vol.~15, no.~2, p.~454, 2023.

\bibitem{azzam2020ten}
S.~I. Azzam, A.~V. Kildishev, R.-M. Ma, C.-Z. Ning, R.~Oulton, V.~M. Shalaev,
  M.~I. Stockman, J.-L. Xu, and X.~Zhang, ``Ten years of spasers and plasmonic
  nanolasers,'' {\em Light: Science \& Applications}, vol.~9, no.~1, p.~90,
  2020.

\bibitem{bordo2017cooperative}
V.~Bordo, ``Cooperative effects in spherical spasers: Ab initio analytical
  model,'' {\em Physical Review B}, vol.~95, no.~23, p.~235412, 2017.

\bibitem{bergman2003surface}
D.~J. Bergman and M.~I. Stockman, ``Surface plasmon amplification by stimulated
  emission of radiation: quantum generation of coherent surface plasmons in
  nanosystems,'' {\em Physical Review Letters}, vol.~90, no.~2, p.~027402,
  2003.

\bibitem{stockman2010spaser}
M.~I. Stockman, ``The spaser as a nanoscale quantum generator and ultrafast
  amplifier,'' {\em Journal of Optics}, vol.~12, no.~2, p.~024004, 2010.

\bibitem{dicke1954coherence}
R.~H. Dicke, ``Coherence in spontaneous radiation processes,'' {\em Physical
  Review}, vol.~93, no.~1, p.~99, 1954.

\bibitem{andreev1980collective}
A.~V. Andreev, V.~I. Emel'yanov, and Y.~A. Il'inski{\u\i}, ``Collective
  spontaneous emission ({Dicke} superradiance),'' {\em Soviet Physics Uspekhi},
  vol.~23, no.~8, p.~493, 1980.

\bibitem{prasad2000polarium}
S.~Prasad and R.~J. Glauber, ``Polarium model: Coherent radiation by a resonant
  medium,'' {\em Physical Review A}, vol.~61, no.~6, p.~063814, 2000.

\bibitem{sivasubramanian2001gauge}
S.~Sivasubramanian, A.~Widom, and Y.~Srivastava, ``{Gauge invariant
  formulations of Dicke-Preparata super-radiant models},'' {\em Physica A:
  Statistical Mechanics and its Applications}, vol.~301, no.~1-4, pp.~241--254,
  2001.

\bibitem{sivasubramanian2001super}
S.~Sivasubramanian, A.~Widom, and Y.~Srivastava, ``Super-radiance and the
  unstable photon oscillator,'' {\em International Journal of Modern Physics
  B}, vol.~15, no.~05, pp.~537--548, 2001.

\bibitem{sivasubramanian2003microscopic}
S.~Sivasubramanian, A.~Widom, and Y.~Srivastava, ``Microscopic basis of thermal
  superradiance,'' {\em Journal of Physics: Condensed Matter}, vol.~15, no.~7,
  p.~1109, 2003.

\bibitem{prasad2010coherent}
S.~Prasad and R.~J. Glauber, ``Coherent radiation by a spherical medium of
  resonant atoms,'' {\em Physical Review A}, vol.~82, no.~6, p.~063805, 2010.

\bibitem{pustovit2009cooperative}
V.~N. Pustovit and T.~V. Shahbazyan, ``Cooperative emission of light by an
  ensemble of dipoles near a metal nanoparticle: The plasmonic {Dicke}
  effect,'' {\em Physical Review Letters}, vol.~102, no.~7, p.~077401, 2009.

\bibitem{pustovit2010plasmon}
V.~N. Pustovit and T.~V. Shahbazyan, ``Plasmon-mediated superradiance near
  metal nanostructures,'' {\em Physical Review B}, vol.~82, no.~7, p.~075429,
  2010.

\bibitem{delga2014quantum}
A.~Delga, J.~Feist, J.~Bravo-Abad, and F.~Garcia-Vidal, ``Quantum emitters near
  a metal nanoparticle: strong coupling and quenching,'' {\em Physical Review
  Letters}, vol.~112, no.~25, p.~253601, 2014.

\bibitem{scheibner2007superradiance}
M.~Scheibner, T.~Schmidt, L.~Worschech, A.~Forchel, G.~Bacher, T.~Passow, and
  D.~Hommel, ``Superradiance of quantum dots,'' {\em Nature Physics}, vol.~3,
  no.~2, pp.~106--110, 2007.

\bibitem{imamog1999quantum}
A.~Imamoglu, D.~D. Awschalom, G.~Burkard, D.~P. DiVincenzo, D.~Loss,
  M.~Sherwin, A.~Small, {\em et~al.}, ``Quantum information processing using
  quantum dot spins and cavity {QED},'' {\em Physical Review Letters}, vol.~83,
  no.~20, p.~4204, 1999.

\bibitem{temnov2005superradiance}
V.~V. Temnov and U.~Woggon, ``Superradiance and subradiance in an
  inhomogeneously broadened ensemble of two-level systems coupled to a low-q
  cavity,'' {\em Physical Review Letters}, vol.~95, no.~24, p.~243602, 2005.

\bibitem{reithmaier2004strong}
J.~P. Reithmaier, G.~Sek, A.~L{\"o}ffler, C.~Hofmann, S.~Kuhn, S.~Reitzenstein,
  L.~Keldysh, V.~Kulakovskii, T.~Reinecke, and A.~Forchel, ``Strong coupling in
  a single quantum dot--semiconductor microcavity system,'' {\em Nature},
  vol.~432, no.~7014, pp.~197--200, 2004.

\bibitem{yoshie2004vacuum}
T.~Yoshie, A.~Scherer, J.~Hendrickson, G.~Khitrova, H.~Gibbs, G.~Rupper,
  C.~Ell, O.~Shchekin, and D.~Deppe, ``Vacuum {Rabi} splitting with a single
  quantum dot in a photonic crystal nanocavity,'' {\em Nature}, vol.~432,
  no.~7014, pp.~200--203, 2004.

\bibitem{greenberg2012steady}
J.~A. Greenberg and D.~J. Gauthier, ``Steady-state, cavityless, multimode
  superradiance in a cold vapor,'' {\em Physical Review A}, vol.~86, no.~1,
  p.~013823, 2012.

\bibitem{burns1993mineralogical}
R.~G. Burns and R.~G. Burns, {\em Mineralogical applications of crystal field
  theory}.
\newblock No.~5, Cambridge University Press, 1993.

\bibitem{faisal1987theory}
F.~Faisal, {\em Theory of multiphoton processes}.
\newblock Springer Science \& Business Media, 1987.

\bibitem{qutip1}
J.~Johansson, P.~Nation, and F.~Nori, ``{QuTiP}: An open-source {P}ython
  framework for the dynamics of open quantum systems,'' {\em Computer Physics
  Communications}, vol.~183, pp.~1760--1772, aug 2012.

\bibitem{qutip2}
J.~Johansson, P.~Nation, and F.~Nori, ``{QuTiP} 2: A {P}ython framework for the
  dynamics of open quantum systems,'' {\em Computer Physics Communications},
  vol.~184, pp.~1234--1240, apr 2013.

\bibitem{prep-weak-coherence}
G.~Preparata, {\em QED coherence in matter}.
\newblock World Scientific Ed., 1995.
\newblock Page 46.

\bibitem{shammah2018open}
N.~Shammah, S.~Ahmed, N.~Lambert, S.~De~Liberato, and F.~Nori, ``Open quantum
  systems with local and collective incoherent processes: Efficient numerical
  simulations using permutational invariance,'' {\em Physical Review A},
  vol.~98, no.~6, p.~063815, 2018.

\bibitem{tindall2020quantum}
J.~Tindall, C.~S. Mu{\~n}oz, B.~Bu{\v{c}}a, and D.~Jaksch, ``Quantum
  synchronisation enabled by dynamical symmetries and dissipation,'' {\em New
  Journal of Physics}, vol.~22, no.~1, p.~013026, 2020.

\end{thebibliography}
\bibliographystyle{ieeetr}

\end{document}